\documentclass[preprint2]{aastex}

\usepackage{psfig}

\newcommand\mzon   {M$_{\odot}$}
\newcommand\pp     {$\pm$}

\def\degr{\hbox{$^\circ$}}
\newcommand\Lunit   {erg s$^{-1}$}
\newcommand\funit   {erg cm$^{-2}$ s$^{-1}$}
\newcommand\nh   {$N_{\rm H}$}

\righthead{A {\itshape Chandra} observation of MXB 1659--29 in
quiescence} \slugcomment{Resubmitted to ApJ: 4 February, 2003}

\begin{document}

\title{A {\itshape Chandra} observation of the neutron star X-ray
transient and eclipsing binary MXB 1659--29 in quiescence}

\author{Rudy Wijnands\altaffilmark{1,2,3}, Mike Nowak\altaffilmark{1},
Jon M. Miller\altaffilmark{4}, Jeroen Homan\altaffilmark{5}, Stefanie
Wachter\altaffilmark{6}, Walter H. G. Lewin\altaffilmark{1} }

\altaffiltext{1}{Center for Space Research, Massachusetts Institute of
Technology, 77 Massachusetts Avenue, Cambridge, MA 02139-4307, USA;
rudy@space.mit.edu}

\altaffiltext{2}{Chandra Fellow}

\altaffiltext{3}{Present address: School of Physics and Astronomy, 
University of St Andrews, North Haugh, St Andrews, Fife KY16 9SS,
Scotland, UK}

\altaffiltext{4}{Harvard-Smithsonian Center for Astrophysics, 
60 Garden Street, Cambridge, MA 02139, USA}

\altaffiltext{5}{Osservatorio Astronomico di Brera, Via E. Bianchi 46,
23807 Merate LC, Italy}

\altaffiltext{6}{Cerro Tololo Inter-American Observatory, National
Optical Astronomy Observatory, Casilla 603, La Serena, Chile}

\begin{abstract}

After almost 2.5 years of actively accreting, the neutron star X-ray
transient and eclipsing binary MXB 1659--29 returned to quiescence in
2001 September.  We report on a {\it Chandra} observation of this
source taken a little over a month after this transition. The source
was detected at an unabsorbed 0.5--10 keV flux of only $(2.7 - 3.6)
\times10^{-13}$ \funit, which implies a 0.5--10 keV X-ray luminosity
of approximately $(3.2 - 4.3) \times10^{33}\,\,\, (d/10\,\,\,{\rm
kpc})^2$ \Lunit, with $d$ the distance to the source in kpc. Its
spectrum had a thermal shape and could be well fitted by either a
blackbody with a temperature $kT$ of $\sim0.3$ keV or a neutron star
atmosphere model with a $kT$ of $\sim0.1$ keV.  The luminosity and
spectral shape of MXB 1659--29 are very similar to those observed of
the other neutron star X-ray transients when they are in their
quiescent state. The source was variable during our observation,
exhibiting a complete eclipse of the inner part of the system by the
companion star. Dipping behavior was observed before the eclipse,
likely due to obscuration by an extended feature in the outer part of
a residual accretion disk.  We discuss our observation in the context
of the cooling neutron star model proposed to explain the quiescent
properties of neutron star X-ray transients.

\end{abstract}


\section{Introduction \label{section:intro}}

During outburst episodes, neutron star X-ray transients can be
detected at luminosities of $\sim10^{36-38}$ \Lunit~(e.g., Chen,
Shrader, \& Livio 1997).  During those outbursts, the transients are
very similar to the persistent sources with respect to their X-ray
properties. The high X-ray luminosity is very likely due to the
accretion of matter onto the neutron star.  These transients are
characterized by their bright outbursts, but most of the time they
spend in a quiescent state in which they are orders of magnitude
dimmer at all wavelengths. Fortunately, using sensitive imaging
instruments, we are still able to detect them at X-ray luminosities of
$\sim 10^{32-34}$ \Lunit~(e.g., van Paradijs et al. 1987; Asai et
al. 1996, 1998). The high sensitivity camera's aboard {\it Chandra}
and {\it XMM-Newton} are well suited to detect quiescent systems and
obtain good X-ray spectra for the brightest systems (see, e.g., Daigne
et al. 2002; in 't Zand et al. 2001; Rutledge et al. 2001a, 2001b;
Wijnands et al. 2001b, 2002b).  To explain the low quiescent X-ray
properties, several models have been developed. For example, the
X-rays could be due to the residual accretion of matter onto the
neutron star or magnetospheric boundary, or the pulsar emission
mechanism might be active (see, e.g., Stella et al. 1994; Corbet 1996;
Campana et al. 1998b; Menou et al. 1999; Campana \& Stella 2000; Menou
\& McClintock 2001). Currently the most successful model is that in
which the X-rays are due to the thermal emission from the neutron star
surface, which will be referred to as 'the cooling neutron star
model'.

\subsection{The cooling neutron star model}

In the cooling neutron star model (e.g., van Paradijs et al. 1987;
Campana et al. 1998b; Brown, Bildsten, \& Rutledge 1998 and references
therein) the emitted radiation below a few keV is thermal emission
originating from the neutron star surface.  Brown et al. (1998) argued
that the neutron star core is heated by the nuclear reactions
occurring deep in the crust when the star is accreting and this heat
is released as thermal emission during quiescence. If the quiescent
emission is dominated by the thermal emission of the cooling neutron
star, then the quiescent luminosity should depend on the time averaged
(over $10^{4-5}$ years) accretion luminosity of the system (Campana et
al. 1998b; Brown et al. 1998).  Thus, the quiescent luminosities of
the detected systems can directly be compared with the predicted ones
obtained from estimates of the long term accretion history of the
sources.

The neutron star cooling model also gives clear predictions for the
spectral shape of the quiescent X-ray spectrum, which should be
thermal.  Although a simple blackbody model can be fitted to the data,
the obtained radii of the emitting regions (of the order of only a few
kilometers) are considerably lower than the predicted radii of neutron
stars (Shapiro \& Teukolsky 1983). To circumvent this discrepancy, it
has been proposed that quiescent neutron star systems do not emit a
true blackbody spectrum but a modified one (Brown et al. 1998). When
using blackbody models to fit such modified spectra, the effective
temperatures will be overestimated and the emitting areas
underestimated. By fitting more realistic models, such as the so-called
'neutron star atmosphere models' (the non-magnetic models are
appropriate for quiescent neutron star systems; e.g., Zavlin, Pavlov,
\& Shibanov 1996 ), to the X-ray data, emitting radii were obtained
which are consistent with the expected radii of neutron stars
(Rutledge et al. 1999, 2000).

The cooling neutron star model cannot fully explain all
characteristics of the quiescent emission. For example, the power-law
shaped spectral component which dominates the quiescent spectra above
a few keV in several systems\footnote{Note that not in all detected
quiescent systems this power-law component could be detected and that
the flux ratio of the power-law component with the thermal component
varies considerably between sources.} (e.g., Asai et al. 1996, 1998;
Campana et al. 1998a) cannot be explained by the cooling models. It is
conceivable that this component might be described by one or more of
the alternative models discussed above (in particular the residual
accretion model). However, the observational results on this component
and our understanding of its nature are very limited.

\subsection{The quasi-persistent X-ray transients}

Recently, a sub-group of neutron star X-ray transients has received
extra attention because of their potential to test the cooling neutron
star model and to determine some of the physical properties of the
neutron star crust and core. These particular transients do not have
traditional outbursts which only last weeks to at most a few months,
but instead they stay active for several years to over a decade (and
maybe even longer). These systems have been called long-duration
transients or quasi-persistent sources (e.g., Wijnands et al. 2001b,
Wijnands 2002). The long outburst behavior of those sources might be
related to the extended episodes (several months to several years) of
low-level activity seen in other transients usually after they have
exhibited bright outbursts (e.g., in 4U 1630--47, Aql X-1, 4U
1608--52, or SAX J1808.4--3658; Kuulkers et al. 1997; Bradt et
al. 2000; Wijnands et al. 2001c; Wachter et al. 2002), although those
episodes are generally less luminous ($<10^{36}$ \Lunit) than the
outbursts of the quasi-persistent sources ($10^{36} - 10^{37}$
\Lunit).

In 'ordinary' (i.e., short-duration) transients, the accretion of
matter will have only a very minor effect on the thermal state of the
crust, but for these quasi-persistent sources the prolonged accretion
episodes can heat the crust to high temperatures, considerably higher
than that of the neutron star core (see Rutledge et al. 2002). When
those systems become quiescent again, it might take years to decades
for the crust to return to thermal equilibrium with the core and the
initial quiescent properties of those systems might be dominated by
the crust emission and not by the state of the core (as is the case in
ordinary transients). Monitoring observations of those systems in
quiescence might even allow one to follow the cooling of the crust
from which the heat conductivity of the crust can be determined
(Rutledge et al. 2002).

Recently, one of the quasi-persistent systems (KS 1731--260) suddenly
turned off after having actively accreted for over 12.5 years. A {\it
Chandra} observation taken a few months after this transition showed
the source at a 0.5--10 keV luminosity of $\sim10^{33}$
\Lunit~(Wijnands et al. 2001b). An {\it XMM-Newton} observation of
this system performed about half a year after the {\it Chandra}
observation, showed that the system had declined by a factor of
$\sim$3 in luminosity (Wijnands et al. 2002b).  If the quiescent
emission from this system was dominated by the state of the crust, the
decrease in luminosity within half a year strongly indicates that the
crust must have a high heat conductivity (Wijnands et al. 2002b; using
the cooling curves calculated for this system by Rutledge et
al. 2002). In this scenario, the core temperature is expected to be
lower than the crust temperature and the luminosity measured with {\it
XMM-Newton} can be used as an upper limit on the core luminosity. The
quiescent luminosity of KS 1731--260 is much lower than expected from
its long term accretion history, and can only be explained in terms of
the standard cooling model if this system is dormant for at least
several thousand of years between outbursts (assuming all outbursts of
this system are very similar to the last one, which might not be a
valid assumption; Wijnands et al. 2001b; Rutledge et
al. 2002). Alternatively, enhanced cooling processes might be active
in the core, rapidly cooling it (e.g., Wijnands et al. 2001b,
2002b).

\subsection {MXB 1659--29}

\begin{figure}[t]
\begin{center}
\begin{tabular}{c}
\psfig{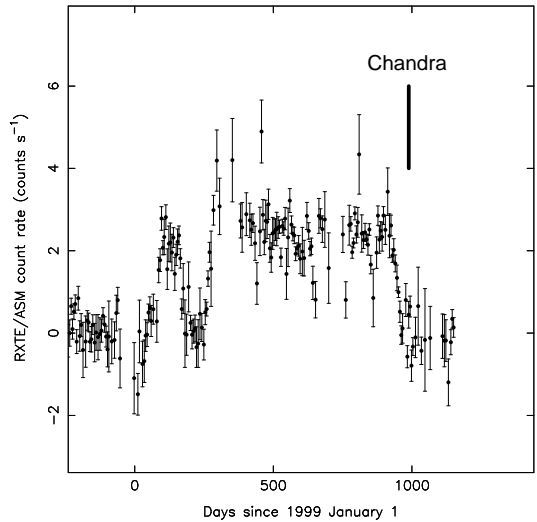}
\end{tabular}
\figcaption{The {\it RXTE}/ASM light curve of MXB 1659--29 clearly
showing the 1999--2001 outburst. The time of our {\it Chandra}
observation is indicated by the solid line. Note that our {\it
Chandra} observation was performed at times that the {\it RXTE}/ASM
(and also the {\it RXTE}/PCA) could not detect the source anymore (see
text).
\label{fig:asm} }
\end{center}
\end{figure}

In 2001 September, the opportunity arose to use another
quasi-persistent system to test the cooling neutron star model: MXB
1659--29.  This source is an X-ray transient and was discovered in
1976 by Lewin, Hoffman, \& Doty (1976) during type-I X-ray bursts,
which clearly demonstrates that the compact object in this system is a
neutron star. The source was detected several times between 1976
October and 1978 September with {\it SAS3} and {\it HEAO} (Lewin et
al. 1978; Share et al. 1978; Griffiths et al. 1978; Cominsky, Ossman,
\& Lewin 1983; Cominsky \& Wood 1984, 1989) and irregular X-ray
variability was found in this system (Lewin 1979; Cominsky et
al. 1983). Cominsky \& Wood (1984, 1989) reported on the discovery of
eclipses every $\sim$7.1 hours, which can be identified with the
orbital period of the system.  MXB 1659--29 is one of only several
non-pulsating neutron star low-mass X-ray binaries for which total
eclipses have been observed (The other confirmed eclipsing neutron
star systems are EXO 0748--676, GRS 1747--312, and AX J1745.6--2901,
although several other systems [e.g., 4U 2129+47] show partial
eclipses; Parmar et al. 1986; In 't Zand et al. 2000; Maeda et
al. 1996; McClintock et al. 1982). During later observations, using a
variety of satellites (e.g., {\it Hakucho}, {\it EXOSAT}, {\it
ROSAT}), the source could not be detected anymore (Cominsky et
al. 1983; Verbunt 2001). The pointed {\it ROSAT} observations in the
early 1990's failed to detected the source with a 0.5--10 keV upper
limit on the unabsorbed flux of $(1 - 2) \times 10^{-14}$
\funit~(Verbunt 2001; Wijnands 2002; Oosterbroek et al. 2001).

The source remained dormant until 1999 April, when in 't Zand et
al. (1999) reported it to be active again in observations obtained
with the {\it BeppoSAX} Wide Field Camera. Wachter, Smale, \& Bailyn
(2000) and Oosterbroek et al. (2001) obtained an updated ephemeris for
the orbital period (using data obtained with the {\it Rossi X-ray
Timing Explorer} [{\it RXTE}] and {\it BeppoSAX}). Studies of its
X-ray spectrum were performed using {\it BeppoSAX} (Oosterbroek et
al. 2001) and {\it XMM-Newton} (Sidoli et al. 2001), and, using {\it
RXTE} data, Wijnands, Strohmayer, \& Franco (2001a) found $\sim$567 Hz
oscillations during X-ray bursts. Those oscillations are likely
related to the neutron star spin frequency. The source remained bright
for almost 2.5 years before it became dormant again in 2001 September
(Wijnands et al. 2002a).  Because of its long outburst duration, MXB
1659--29 may be classified as a quasi-persistent source. We had a
Cycle 3 {\it Chandra} TOO proposal approved to obtain a quiescent
observation of the next quasi-persistent source that could turn off,
within a month after the transition. As part of this proposal, MXB
1659--29 was observed on 2001 October 15 using {\it Chandra} for
$\sim19$ ksec.  Here we report on this observation.

\section{Observation, analysis, and results}

\begin{figure}[t]
\begin{center}
\begin{tabular}{c}
\psfig{figure=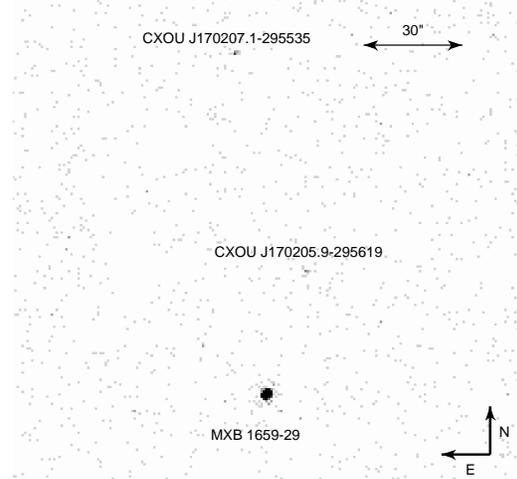,width=7cm}
\end{tabular}
\figcaption{The {\it Chandra}/ACIS-S 0.5--8 keV image of the field of
MXB 1659--29. The source is clearly detected together with two other
weak X-ray sources.
\label{fig:image} }
\end{center}
\end{figure}

The {\it RXTE} All Sky Monitor (ASM) light curve of MXB 1659--29 is
shown in Figure~\ref{fig:asm} (see also Wijnands et al. 2002a). As can
clearly be seen, MXB 1659--29 stayed active during its last outburst
for almost 2.5 years before it could not be detected anymore with the
{\it RXTE}/ASM at the end of 2001 August. During more sensitive
observations using the proportional counter array (PCA) aboard {\it
RXTE}, the source could still be detected until 2001 September 7, but
it was undetectable in observations performed on 2001 September 14,
24, and 30 (Wijnands et al. 2002a; with upper limits on the flux of
0.5--1 mCrab; 2--60 keV).

After the non-detection using the {\it RXTE}/PCA on 2001 September 14,
we triggered our Cycle 3 {\it Chandra} proposal and a {\it
Chandra}/ACIS-S observation on MXB 1659--29 was performed on 15
October 2001 between 16:31 and 22:24 UTC for a total of $\sim$18.8
ksec of on-source time. No background flares occurred during the
observation so all data could be used.  We used the CIAO tools
(version 2.2.1) and the standard {\it Chandra} analysis
threads\footnote{Listed at http://asc.harvard.edu} to analyze the
data. During our observation the ACIS-S3 CCD was used with a 1/4
sub-array (resulting in a frame time of 0.8 seconds). This
configuration was used to reduce the possible pile-up problems in case
the source had exceeded a flux level of $\sim10^{-12}$ \funit. As
demonstrated below, only about 2\% pile-up occurred during our
observation because of the relatively low flux of the source.

\subsection{X-ray image and light curve of MXB 1659--29 \label{section:image}}

\begin{figure}[t]
\begin{center}
\begin{tabular}{c}
\psfig{figure=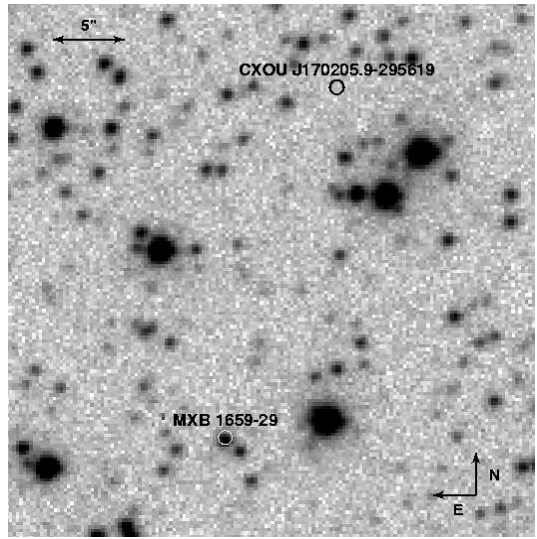,width=7cm}
\end{tabular}
\figcaption{The optical I band image from Wachter et al. (2000) during
outburst together with the {\it Chandra} positions of MXB 1659--29 and
CXOU J170205.9--295619, using circles with radii of 0.5$''$. Clearly,
the {\it Chandra} position of MXB 1659--29 is fully consistent with
the optical one.
\label{fig:optical} }
\end{center}
\end{figure}

The obtained 0.5--8 keV image of the region around MXB 1659--29 is
shown in Figure~\ref{fig:image}. We used the tool {\it wavdetect} to
search the complete chip for point sources. The three sources visible
in Figure~\ref{fig:image} are detected (their coordinates are listed
in Tab.~\ref{tab:sources}) together with two additional, rather
diffuse sources, which could be truly diffuse sources or due to an
elevated background count rate at those positions. In this paper, we
are only interested in the properties of MXB 1659--29 so we do not
further investigate the significance of those potential sources. We
used the optical I-band images of Wachter et al. (2000) taken during
outburst to determine possible optical counterparts for the detected
{\it Chandra} sources. We tied these optical images to the 2MASS J
image available for this region; the resulting image, including the
{\it Chandra} positional circles, is shown in
Figure~\ref{fig:optical}. Clearly, the {\it Chandra} position of MXB
1659--29 is consistent with its optical position during outburst, with
an offset of only $\sim0.15''$, well within the bore sight error
(Aldcroft et al. 2000) and the errors on the optical position.  The
positions listed in Table~\ref{tab:sources} are adjusted for this
small off-set.  The two extra {\it Chandra} sources do not have
counterparts in the optical image.

Figure~\ref{fig:eclipse} displays the {\it Chandra}/ACIS-S light
curve. It clearly shows an eclipse, which is very likely due to the
obscuration of the inner X-ray emitting region by the companion star
(cycle 30823 using the linear ephemeris of Oosterbroek et
al. 2001). The eclipse duration (842\pp90 seconds) and the egress and
ingress times ($<247$ and $< 400$ seconds, respectively) have been
calculated using the method outlined by Nowak et al. (2002). The
obtained eclipse duration is consistent with that observed during
outburst ($\sim900$ seconds; Wachter et al. 2000). Eclipses in
quiescence have also been observed for the neutron star transient 4U
2129+47 (Nowak et al. 2002) and from the likely neutron star transient
X5 in the globular cluster 47 Tuc (Heinke et al. 2001, 2002).
Besides the eclipse, the {\it Chandra}/ACIS-S light curve also shows
clear dipping behavior a few kiloseconds before the eclipse, which
again is very similar to 4U 2129+47 and X5 in 47 Tuc.  Those dips
might be due to the same process causing the dipping behavior of the
source when it is still actively accreting: the dips might be due to
obscuration of the central X-ray source by a large structure in the
outer accretion disk, possible the impact point of the accretion
stream from the companion star and the accretion disk (see also Nowak
et al. 2002 and Heinke et al. 2001, 2002 for a discussion about the
dips for 4U 2129+47 and X5, respectively).  The decrease in count rate
would then be due to an increase in the absorption column in front of
the central X-ray source. This hypothesis can be tested by examining
the spectrum of the source in and outside the dip (see
\S~\ref{section:dipspectrum}). 
Note that this hypothesis requires that residual disks are still
present in quiescence without accretion occurring onto the neutron
stars. Indeed such residual disks are expected from the disk
instability model to explain X-ray outbursts (see Lasota 2001 for a
review) and independent proof for such disks came from optical
observations from quiescent X-ray transients (e.g., McClintock
\& Remillard 2000).

\subsection{The X-ray spectrum of the persistent emission}

\begin{figure}[t]
\begin{center}
\begin{tabular}{c}
\psfig{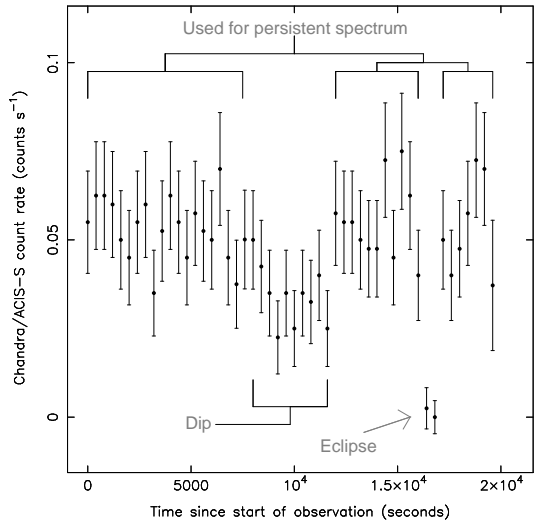}
\end{tabular}
\figcaption{The {\it Chandra}/ACIS-S 0.5--8 keV light curve of MXB
1659--29. Clearly visible are the eclipse and dipping behavior (about
6 ksec before the eclipse). It is indicated in the figure which part
of the data has been used to extract the persistent spectrum of MXB
1659--29 (see Fig.~\ref{fig:spectrum}) and the dip spectrum. Each
point represents 400 seconds of data.
\label{fig:eclipse} }
\end{center}
\end{figure}

All X-ray spectra were extracted using a circle with $5''$ in radius
on the position of the source as determined using {\it wavdetect} and
the background data were obtained by using an annulus on the source
position with an inner radius of $7''$ and an outer one of $20''$. All
obtained spectra were rebinned using the FTOOLS tool {\it grppha} into
bins with a minimum of 15 counts per bin and we fitted the spectrum
using XSPEC (version 11.1; Arnaud 1996). We extracted spectra for the
part of the observation during which the source was relatively stable
(called the 'persistent spectrum'; see Fig.~\ref{fig:eclipse};
$\sim13.2$ ksec of data) and for the part during the dips (called the
'dip spectrum'; Fig.~\ref{fig:eclipse}; a total of $\sim$3.8 ksec of
data).

Many single component models fit the persistent spectrum well;
however, the quiescent spectra of neutron star X-ray transients are
most often fit with a blackbody model or a neutron star atmosphere
model. Therefore, we concentrate on those two models, using the model
from Zavlin et al. 1996 (the non-magnetic case) as the atmosphere
model.  Theoretically it is expected that the emerging spectrum should
resemble that assumed by the atmosphere model (e.g., Brown et
al. 1998), however, observationally both the atmosphere model and the
black body model produce equally satisfactory fits to the data.
Because also our data do not allow for a rejection of the blackbody
model, and to allow comparison with previous results on the quiescent
spectra of neutron star X-ray transients, we also report the blackbody
results.  In certain systems, a power-law tail above a few keV was
found, and although such a power-law component was not required by the
data, we fitted the spectra with the above two models including a
power-law component with a photon index of 1 or 2 to obtain an upper
limit on this component.

\begin{figure}[t]
\begin{center}
\begin{tabular}{c}
\psfig{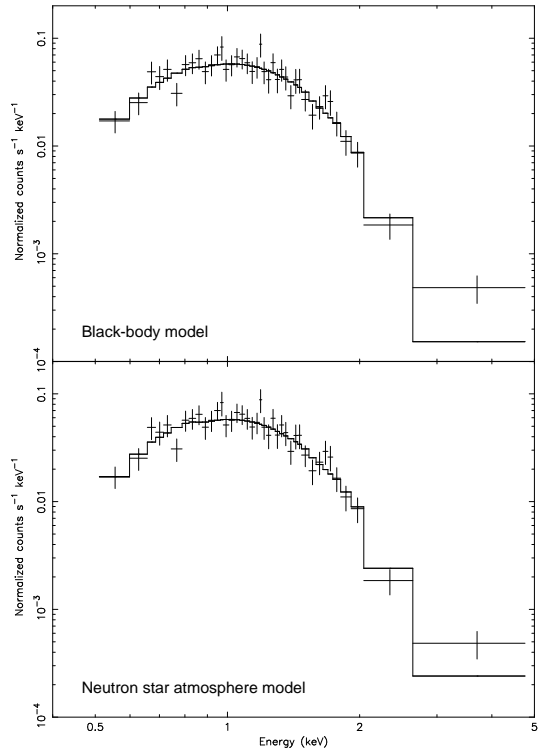}
\end{tabular}
\figcaption{The {\it Chandra}/ACIS-S spectrum of MXB 1659--29 of the
persistent part of the data. In the top panel, the solid line
represents the best blackbody fit to the data, in the bottom panel the
best neutron star atmosphere model fit.
\label{fig:spectrum} }
\end{center}
\end{figure}

The spectral results are listed in Table~\ref{tab:spectrum} and the
persistent spectrum is shown in Figure~\ref{fig:spectrum}.  The column
density \nh~was left free in the fits and the obtained value is in the
range of what has been observed previously for this source during
outburst using {\it BeppoSAX} and {\it XMM-Newton} ($0.13-0.35
\times10^{22}$ cm$^{-2}$; Oosterbroek et al. 2001; Sidoli et
al. 2001). Both the blackbody and the atmosphere model fit the data
well and the best-fit temperature was $\sim$0.3 keV for the blackbody
fits and $\sim$0.1 keV for the atmosphere model\footnote{Using the
obtained spectral parameters in this paragraph and the 0.8 frame time
of our observation, we have used PIMMS to estimate the degree of
pile-up in our spectrum. We found that about 2\% of the photons should
be piled-up and the effects on the X-ray spectral parameters were only
marginal. Therefore, we have not corrected the X-ray spectrum for this
effect.}.  When assuming a distance of 10 kpc\footnote{The distance
toward the source is not well known. Using the luminosity of the X-ray
bursts detected during the 1999--2001 outburst, Oosterbroek et
al. (2001) obtained a range of 11--13 kpc but Muno et al. (2001)
reported 10 kpc. In the remaining we will assume 10 kpc, but we note
that it is rather uncertain. When appropriate we will discuss the
effect of this uncertainty on the interpretation of our results.}, the
radius for the emitting area was only a few kilometers using the
blackbody model which is lower than the theoretical expected radius
for a neutron star. The atmosphere model gave radii that were
consistent with those expected if the emission arose from the neutron
star and the complete surface was radiating, favoring the atmosphere
model. For larger assumed distances the radii would increase, although
the distance has to be unrealistically large ($>20$ kpc) for the
radius obtained via the black-body model to become consistent with
theoretical expectations. The atmosphere model does not allow for the
distance to become much larger than assumed because the obtained
radius will quickly be inconsistent with theoretical expectations.

The obtained unabsorbed 0.5--10 keV fluxes were between 3.0 and 3.6
$\times 10^{-13}$ \funit, resulting in a luminosity of $(3.6 - 4.3)
\times10^{33}\,\,\, (d/10\,\,\,{\rm kpc})^2$ \Lunit. When including a
power-law component in the fit, it could not be detected significantly
and its 0.5--10 keV flux could be constrained to be less than about
25\%--35\% (depending on photon index) of the 0.5--10 keV blackbody
flux or that of the atmosphere component.

\begin{figure}[t]
\begin{center}
\begin{tabular}{c}
\psfig{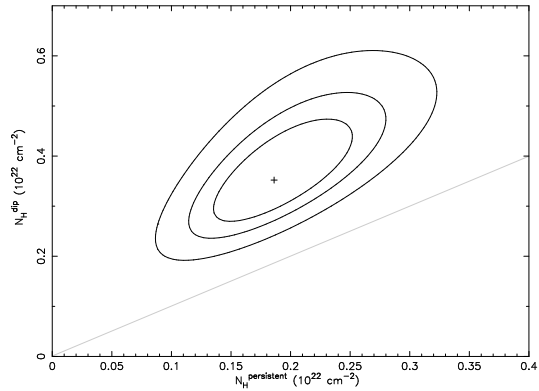}
\end{tabular}
\figcaption{The error ellipse of the fitted column density of the
persistent emission part of the data ($N_{\rm H}^{\rm persistent}$)
versus that of the dip part ($N_{\rm H}^{\rm dip}$). The model used in
XSPEC was the bbodyrad model. The plus represents the best fit value
and the ellipses are for 1, 2, and 3$\sigma$ confidence levels (for
two parameters). The straight line represents $N_{\rm H}^{\rm dip}=
N_{\rm H}^{\rm persistent}$.
\label{fig:nh} }
\end{center}
\end{figure}

\subsection{The spectrum during the dip \label{section:dipspectrum}}

Due to the low statistics of the dip data, the source spectrum during
the dip could not be constrained by using only the dip data. However,
by assuming that the decrease in count rate is only due to an increase
in absorption by obscuring material (as has been found during
outburst; e.g., Oosterbroek et al. 2001; Sidoli et al. 2001), we
fitted the persistent spectrum and the dip spectrum simultaneously
using the same blackbody or atmosphere model but with different column
densities. The spectral results are also listed in
Table~\ref{tab:spectrum}. The spectral parameters of the persistent
emission were consistent with those obtained when only fitting the
persistent emission spectrum, although the measured fluxes tend to be
slightly lower. The column density of the dip spectrum is higher (by a
factor of two) than that obtained for the persistent
emission. Figure~\ref{fig:nh} shows the error ellipse of the column
density of the persistent emission spectrum versus that of the dip
spectrum. Clearly, at high significance it is rejected that the column
density of the dip spectrum is identical to that of the persistent
emission. A systematic trend is clearly seen that the dip column
density is always larger than that of the persistent emission (a
similar result is obtained when the atmosphere model is used instead
of the blackbody model).  Our results show that our data are
consistent with a scenario in which the intrinsic source spectrum
remains identical throughout the {\it Chandra} observation, and that
the decrease in count rate during the dip is due to an increase in
absorption, but we cannot exclude other mechanisms responsible for the
count rate decrease. For example, the data are also consistent with a
decrease in luminosity without a change in spectral shape or with a
decrease of the temperature.

The observation of a dip in the quiescent X-ray light curve of MXB
1659--29 might imply that absorption by the outer disk also could
occur, albeit less strong, during the orbital phases where no dips are
observe, possibly effecting the spectral fits of the persistent
emission. However, during the outburst phase of the source type-I
X-ray bursts were observed and nearly-coherent oscillations during
those bursts were seen (Wijnands et al. 2001a), implying that the
neutron star is observed directly. Very likely, this will also be true
during quiescence.  Furthermore, the measured column density toward
the source from the persistent quiescent emission is rather low and
consistent with other estimates and measurements (Dickey \& Lockman
1990; Oosterbroek et al. 2001; Sidoli et al. 2001) making additional
absorption by the outer disk also unlikely. Therefore, we conclude
that very likely we see the neutron star directly at orbital phases
outside the dip and the eclipse, and our spectral results can directly
be compared with the results obtained from other quiescent neutron
star transients which have lower binary inclination than MXB 1659--29.

\section{Discussion\label{section:discussion}}

We have presented a {\it Chandra} observation of MXB 1659--29
performed $\sim$5 weeks after the last clear detection of the source
with the {\it RXTE}/PCA (indicating that at that time the source was
still actively accreting; Wijnands et al. 2002a). During our {\it
Chandra} observation, we detected the source at a luminosity of $(3.2
- 4.3) \times10^{33} (d/10\,\,\, {\rm kpc})^2$ \Lunit, and its
spectrum could be well described by a thermal component (either a
blackbody model or a neutron star atmosphere model).  The obtained
luminosity and the shape of the X-ray spectrum of MXB 1659--29
resemble those obtained for other quiescent neutron star systems,
strongly suggesting that the source was quiescent during our {\it
Chandra} observation.

\subsection{The cooling neutron star model}

As argued by Rutledge et al. (2002), for systems which are actively
accreting for long periods, the crust might have been heated to very
high temperatures (probably considerably higher than that of the core)
and the quiescent properties might be dominated by the thermal state
of the crust rather than that of the core. Although the duration of
the accretion episode of MXB 1659--29 is considerably shorter (factor
of $\sim$5) than that of KS 1731--260 or X 1732--304 (both of which
had accretion episodes of more than a decade; Wijnands et al. 2001b;
Wijnands et al. 2002b; see also Guainazzi, Parmar, \& Oosterbroek 1999
for X 1732--304), here we assume that the 2.5 year accretion episode
of MXB 1659--29 has had a considerable effect on the state of the
crust, similar to what has been argued for KS 1731--260 (Rutledge et
al. 2002; although likely less extreme), .

This assumption is also supported by the fact that in the early
1990's, the source could not be detected in quiescence using a {\it
ROSAT} observation and only a 0.5--10 keV upper limit of $(1 - 2)
\times 10^{32} (d/10\,\,\, {\rm kpc})^2$ \Lunit~ could be obtained
(Verbunt 2001; Wijnands 2002; Oosterbroek et al. 2001) which is about
an order of magnitude lower than the luminosity we have detected
during our {\it Chandra} observation. Therefore, we can conclude that
the {\it Chandra} quiescent luminosity is not the rock bottom
quiescent luminosity of this system and the quiescent properties for
MXB 1659--29 during the {\it Chandra} observation were dominated by
that of the crust and not by the core. As an upper limit on the flux
due to the cooling neutron star core, we will assume the upper limit
provided by {\it ROSAT} and this upper limit will be used to test the
cooling neutron star model.

In order to test the model, the time averaged accretion rate has to be
estimated. The last outburst was fully covered with the {\it RXTE}/ASM
instrument (Fig.~\ref{fig:asm}) and lasted for $\sim2.5$ years. The
2--10 keV luminosity during this state as obtained with {\it BeppoSAX}
and {\it XMM-Newton} was about $6\times 10^{-10}$ \funit~(Oosterbroek
et al. 2001; Sidoli et al. 2001). Using the spectral model and the
spectral parameters given by Oosterbroek et al. (2001), we inferred
(by simulating the spectrum in XSPEC) that the bolometric luminosity
can be at least a factor of two larger. However, the luminosity itself
was variable (by a factor of a few; Fig.~\ref{fig:asm}) during the
outburst, and large uncertainties might be present in the 'typical'
outburst bolometric luminosity. However, in the rest of the discussion
we assume that bolometric outburst flux was typically $(5 -10) \times
10^{-10}$ \funit~during the last outburst.

The 1999--2001 outburst could be an atypical one for MXB 1659--29 and
previous outbursts might have been less bright and/or less long (i.e.,
more like those of the ordinary transients). We have searched the
literature for reports on detections of this source in the past and we
found that the source was conclusively detected in X-rays in 1976
October, 1977 June, July\footnote{During 1977 June and July, the
observations were not very sensitive to persistent X-ray emission and
none was detected (e.g., Cominsky \& Wood 1984, 1989). But X-ray
bursts were observed from the source, indicating that also during
those observations, the source was accreting, albeit at a low level.},
and September, and 1978 March and September using {\it SAS3} and {\it
HEAO} (Lewin et al. 1976; Lewin et al. 1978; Share et al. 1978;
Griffiths et al. 1978; Lewin 1979; Cominsky et al. 1983; Cominsky \&
Wood 1984, 1989) and in the optical on 1978 June 1 and 1979 June 27 --
July 2 (Doxsey et al. 1979; Cominsky et al. 1983). No information is
available during the periods in-between those observations. Although
it cannot be excluded that at those occasions, the source was in
quiescence, we consider it unlikely that the source would only be
active during times when it was observed with an X-ray or optical
instrument and dormant when no instrument looked at the
source. Therefore, it is likely that during the complete period from
1976 October until early July 1979 the source was actively accreting
for over 2.5 years, especially because the recent outburst had a
similar duration. If true, then this would constitute the first
indication that different outbursts of quasi-persistent sources may
have similar durations and that the long duration of those outbursts
might be a common property of those sources.

The first reported non-detection of the source was on 1979 July 17--25
(Cominsky et al. 1983) in optical (V$>$22--23) and with {\it Hakucho}
(no X-ray upper limits were provided). Prior to 1976, the source might
have also been detected during the period 1971 to 1973 using {\it
Uhuru} (classified as 4U 1704--30; Forman et al. 1978), although this
identification with MXB 1659--29 is not certain and we will assume
that they are two different objects (if 4U 1704--30 can be identified
with MXB 1659--29 then the source might have exhibited an extra
outburst during that period or it might have been active for a period
of $\sim$7 years).  The exact fluxes during the observations in the
period October 1976 to early July 1979 are in the range 1 to 6 $\times
10^{-10}$ \funit, but they have large uncertainties because the exact
energy range was not always quoted (if quoted it was 1--10 keV or
2--10 keV), it was unclear if the fluxes were absorbed or unabsorbed,
and the assumed spectral shape was not always similar (often assumed
to be Crab like) and quite different than observed with {\it BeppoSAX}
and {\it XMM-Newton} during the 1999--2001 outburst (Oosterbroek et
al. 2001; Sidoli et al. 2001). However, we will assume that all fluxes
are for the 2--10 keV range and unabsorbed, and a bolometric flux of
about twice the quoted values (as inferred above for the {\it
BeppoSAX} results on MXB 1659--29). Therefore, during the outburst in
the late 1970's, the source was actively accreting for a period of at
least $\sim$2.5 years at a bolometric flux level of 2 to 12 $\times
10^{10}$
\funit. Although quite uncertain, this is remarkably similar to the
values of the 1999--2001 outburst and for simplicity we assume that
the typical outburst duration is 2.5 years and that the bolometric
fluxes during outburst is 5--10 $\times 10^{-10}$ \funit.

Using the Brown et al. (1998) model (assuming standard cooling
processes), the predicted quiescent flux $ F_{\rm q}$ for this source
would then be (Wijnands et al. 2001b; see also Rutledge et al. 2002)
$ F_{\rm q} \approx {t_{\rm o} \over t_{\rm o} + t_{\rm q}} \times
{\langle F_{\rm o} \rangle \over 135}$, with $\langle F_{\rm o}
\rangle$ the average flux during outburst (5--10 $\times 10^{-10}$
\funit), $t_{\rm o}$ the average time the source is in outburst (2.5
years), and $t_{\rm q}$ the average time the source is in quiescence
($\sim$21 year). This results in a predicted quiescent flux of 4--8
$\times 10^{-13}$ \funit. Remarkably, this value is very similar to
the quiescent flux observed during our {\it Chandra}
observations. However, as explained above, based on the {\it ROSAT}
non-detection of the source, the core flux is likely at least an order
of magnitude lower than this, which would make the predicted core flux
considerably higher than that truly originating from the core.
Similar to KS 1731--260 (Wijnands et al. 2001b, 2002b) this low core
flux (and thus temperature) might be due to enhanced core cooling
instead of the assumed standard core cooling in the Brown et
al. (1998) model.

Despite the fact that the last two outburst episodes are likely of
similar duration, it cannot be excluded that they are not typical for
the source and that most of the time MXB 1659--29 exhibits short
duration outbursts. If the typical outburst duration of MXB 1659--29
is not 2.5 years but instead 0.25 years (3 months) or shorter, then
the predicted quiescent flux will be consistent (within the
uncertainties of the model and assumptions) with the {\it ROSAT} upper
limit. The higher {\it Chandra} quiescent luminosity is again due to
the state of the crust which should be considerably heated during the
long accretion episode.

\subsection{Crust cooling}

If during the {\it Chandra} observation the X-ray emission was
dominated by thermal emission from the crust, then further quiescent
observations of MXB 1659--29 will enable studies of the cooling of
the crust in this system. The {\it ROSAT} flux upper limit suggests
that the crust flux should eventually decrease to at least this
level. When the crust will be thermally relaxed with the core, no
significant further decrease of the quiescent flux is expected and
from this bottom flux level the state of the core can be inferred from
which the cooling models can be better constrained. For KS 1731--260
it had already been found that its quiescent luminosity decreased by a
factor of 3 within half a year time (likely due to a temperature
decrease), indicating a highly conductive crust (Wijnands et
al. 2002b). It would be of interest to determine if such a rapid
cooling will also be observed for MXB 1659--29 or if the neutron star
crust in this system has a significantly lower conductivity.

In the latter case, the system should be at the {\it Chandra}
quiescent luminosity for several years to decades. Using the fact that
13 years after the 1976--1978 outburst {\it ROSAT} observed an order
of magnitude lower quiescent luminosity we might set already an upper
limit on the crust cooling time.  When assuming that shortly after the
end of the 1976--1978 outburst the quiescent luminosity was similar to
our measured {\it Chandra} luminosity, the cooling time of the crust
is at least about a factor of 10 in luminosity per decade. Due to
differences in quiescent times, outburst times, and the time averaged
accretion rates between MXB 1659--29 and KS 1731--260 the cooling
curves calculated for KS 1731--260 by Rutledge et al. (2002) cannot be
used for MXB 1659--29. However, if those MXB 1659--29 cooling curves
resemble those of KS 1731--260, then tentatively it might be concluded
that also the neutron star crust in MXB 1659--29 has a high
conductivity (and enhanced core cooling is suggested). We await
specifically calculated cooling curves for MXB 1659--29 and further
monitoring observations using {\it Chandra} or {\it XMM-Newton} in
order to be conclusive about the properties of the neutron star in
this system.

\subsection{Very low thermal emission in quiescent neutron star
systems?}

For most quiescent neutron star X-ray transients it has been inferred
that the thermal emission is in the range of a few times $10^{32-33}$
\Lunit. However, recently, indications have been found (using {\it
XMM-Newton} data) that the thermal emission from the accretion driven
millisecond X-ray pulsar SAX J1808.4--3658 in its quiescent state
might be as low as a few times $10^{30}$ \Lunit~(Campana et al. 2002;
who suggested enhanced core cooling for this low thermal
luminosity). Although the statistics of those results were not
overwhelming and have to be confirmed with additional observations, it
is an interesting possibility and if the processes which produce the
quiescent emission (which has a power-law shape; Campana et al. 2002)
for this system would become inactive, SAX J1808.4--3658 would become
rather dim in quiescence.  Such weak quiescent neutron star systems
might also be suggested by the indications found for enhanced cooling
processes in certain systems, such as MXB 1659--29. The possibility of
dim quiescent neutron star systems raises the question of how dim
certain neutron star systems can become?

The answers to this question will have implications for our
understanding of quiescent X-ray binaries. It has been found that
those X-ray binaries which harbor a black hole instead of a neutron
star can be at least one to two orders of magnitude less luminous in
quiescence than the average neutron star system. This difference has
been used as evidence that the black holes have event horizons (e.g.,
Garcia et al. 2001 and references therein) in contrast to the surfaces
of neutron stars. However, if certain neutron star systems might
become similarly dim (e.g., due to enhanced core cooling or very dim
outbursts), than this luminosity difference will disappear and with it
the evidence for event horizons in black hole systems. The quiescent
spectrum of SAX J1808.4--3658 (Campana et al. 2002) indicates that the
spectrum of such systems might not be dominated by a thermal component
but might have a power-law shape, similar to what has seen for the
black hole systems (Kong et al. 2002) removing also the spectral
differences.

Another area in which it might be important to determine the full
range of the luminosity distribution of quiescent neutron star
systems, is that of the study of the low-luminosity X-ray sources in
globular clusters. Based on their luminosities and their X-ray spectra
(when enough statistics are available), those sources which have
luminosities above a few times $10^{32}$ \Lunit, have been classified
as quiescent neutron star systems and those which have luminosities
below $10^{32}$ \Lunit, as another type of object (possible
cataclysmic variables or millisecond radio pulsars). However, the
observed low luminosity of $5 \times 10^{31}$ \Lunit~ of SAX
J1808.4--3658 (Dotani, Asai, \& Wijnands 2000; Campana et al. 2002)
already shows (irrespective of what exactly causes the X-rays in this
source) that those dim sources might be quiescent neutron star
transients. The fact that the spectrum of SAX J1808.4--3658 appears to
be considerably harder (Campana et al. 2002) than the average
quiescent neutron star spectrum indicates that classifications on
hardness ratio might lead to erroneous results. The possibility that
the thermal emission of SAX J1808.4--3658 might be very low even
suggests that those sources which have luminosities down to only a few
times $10^{30}$ \Lunit~might also be quiescent neutron star systems.
Therefore, we conclude that classifying low-luminosity globular
cluster sources based on their luminosity and broad band spectral
shape (i.e., hardness ratio's) might possibly lead to misleading
results. The full extent of those errors depends on the full
luminosity distribution of the quiescent neutron star transients and
how the spectrum correlates with the luminosity. Those properties have
to be determined before a complete picture of the nature of the
low-luminosity globular cluster sources can be
understood\footnote{Although we have only discussed quiescent neutron
star systems, similar uncertainties in the luminosity distribution of
CVs or millisecond radio pulsars exist. Those uncertainties in the
properties of those systems have to be resolved before a full
understanding of the low-luminosity globular cluster source population
can be obtained.}.

\acknowledgments

RW was supported by NASA through Chandra Postdoctoral Fellowship grant
number PF9-10010 awarded by CXC, which is operated by SAO for NASA
under contract NAS8-39073. This research has made use of the
quick-look results provided by the {\it RXTE}/ASM team, and the data
and resources obtained through the HEASARC online service, provided by
NASA-GSFC. This publication makes use of data products from the Two
Micron All Sky Survey, which is a joint project of the University of
Massachusetts and the Infrared Processing and Analysis
Center/California Institute of Technology, funded by the National
Aeronautics and Space Administration and the National Science
Foundation.

\begin{deluxetable}{lcc}
\tablecolumns{3}
\tablewidth{0pt} 
\tablecaption{Measured {\it Chandra} positions of the detected sources\label{tab:sources}}
\tablehead{
Name                   & \multicolumn{2}{c}{Coordinates}                                               \\
                       &  R.A.                                 & Dec.                                   }
\startdata
MXB 1659--29           & 17$^h$ 02$^m$ 06.54$^s$ \pp0.02$^s$ & --29\degr~ 56$'$ 44.1$''$ \pp0.3$''$ \\
CXOU J170207.1--295535 & 17$^h$ 02$^m$ 07.06$^s$ \pp0.02$^s$   & --29\degr~ 55$'$ 34.7$''$ \pp0.3$''$   \\
CXOU J170205.9--295619 & 17$^h$ 02$^m$ 05.93$^s$ \pp0.02$^s$   & --29\degr~ 56$'$ 19.2$''$ \pp0.4$''$   \\
\enddata
 
\end{deluxetable}

\begin{deluxetable}{lcccc}
\tablecolumns{5}
\tablewidth{0pt} 
\tablecaption{Spectral results\label{tab:spectrum}}
\tablehead{
Parameter& \multicolumn{2}{c}{Persistent} & \multicolumn{2}{c}{Persistent + dip}\\
         &      Blackbody & Hydrogen atmosphere & Blackbody & Hydrogen atmosphere}
\startdata
$N_{\rm H}^{\rm persistent}$ ($10^{22}$ cm$^{-2}$) & $0.22^{+0.07}_{-0.06}$  & $0.28^{+0.13}_{-0.06}$ & $0.19^{+0.07}_{-0.06}$     & 0.27\pp0.07            \\                       
$N_{\rm H}^{\rm dip}$ ($10^{22}$ cm$^{-2}$)        & --                      & --                     & $0.35^{+0.13}_{-0.08}$     & $0.44^{+0.13}_{-0.10}$ \\
$kT$ (keV)                                         & 0.28\pp0.02             & $0.10^{+0.04}_{-0.03}$ & $0.28^{+0.02}_{-0.03}$     & $0.11^{+0.08}_{-0.05}$ \\   
Radius  (${d\over10~{\rm kpc}}$ km)          & $2.4^{+1.5}_{-0.8} $    & $17^{+19}_{-4}$        & $2.1^{+1.6}_{-0.7}$        & $14^{+26}_{-9}$        \\
$F_{\rm persistent}$                               & 3.0                     & 3.6                    & 2.7                        & 3.4                    \\                               
$\chi^2$/dof                                       & 32.2/37                 & 31.0/36                & 45.2/43                    & 42.6/42                \\                               
\enddata
 
\tablenotetext{\,}{Note: The error bars represent 90\% confidence
levels. The fluxes are unabsorbed, in the 0.5--10 keV range, and in
units of $10^{-13}$ \funit. As blackbody model, we used the bbodyrad
model in XSPEC. For the hydrogen atmosphere model, we used the model
of Zavlin et al. 1996 (the non-magnetic case) and for this model the
neutron star mass was fixed to 1.4 \mzon~and the temperature and
radius are for an observer at infinity. For both models the distance
$d$ was assumed to be 10 kpc.}

\end{deluxetable}

\end{document}